\documentclass[aip,apl,a4,reprint,twoside,floatfix,superscriptaddress]{revtex4-1} 

\usepackage{graphicx}
\usepackage{amsmath}
\usepackage{amssymb}
\usepackage{amsfonts}
\usepackage{dcolumn}
\usepackage{epsfig}
\usepackage{subfigure}
\usepackage{bm}

\begin{document}

\title{\bf Spontaneous exchange bias in a nanocomposite of BiFeO$_3$-Bi$_2$Fe$_4$O$_9$}

\author {Tuhin Maity} \affiliation {Micropower-Nanomagnetics Group, Microsystems Center, Tyndall National Institute, University College Cork, Lee Maltings, Dyke Parade, Cork, Ireland}
\author {Sudipta Goswami} \affiliation {Nanostructured Materials Division, CSIR-Central Glass and Ceramic Research Institute, Kolkata 700032, India}
\author {Dipten Bhattacharya} \affiliation {Nanostructured Materials Division, CSIR-Central Glass and Ceramic Research Institute, Kolkata 700032, India}
\author {Gopes C. Das} \affiliation {School of Materials Science, Jadavpur University, Kolkata 700032, India}
\author {Saibal Roy}
\email{saibal.roy@tyndall.ie} \affiliation {Micropower-Nanomagnetics Group, Microsystems Center, Tyndall National Institute, University College Cork, Lee Maltings, Dyke Parade, Cork, Ireland}

\date{\today}

\begin{abstract} 
We have observed a large as well as path-dependent spontaneous exchange bias (H$_{SEB}$) ($\sim$30-60 mT) in a nanocomposite of BiFeO$_3$-Bi$_2$Fe$_4$O$_9$ across 5-300 K when it is measured in an unmagnetized state following zero-field cooling and appropriate demagnetization. The path dependency yields a variation in the exchange bias depending on the sign of the starting field and the path followed in tracing the hysteresis loop. The asymmetry thus observed - $\Delta$H$_{SEB}$ - is found to be decreasing nonmonotonically across 5-300 K with a peak around 200 K. The $\Delta$H$_{SEB}$ together with large H$_{SEB}$ could have significant ramification in tuning the exchange bias driven effects and consequent applications. 
\end{abstract}
\maketitle

While the conventional exchange bias (CEB) is observed under field cooling, which sets the unidirectional anisotropy across a ferromagnet (FM) - antiferromagnet (AFM) interface prior to the measurement of the hysteresis loop, the spontaneous one is observed even in an unmagnetized state following zero-field cooling. It results from a symmetry breaking across the FM-AFM interfaces and setting of the  unidirectional anisotropy (UA) under the first field of the hysteresis loop tracing. In recent times, the spontaneous exchange bias (SEB) has been reported for different alloy and nanoparticle composite systems.\cite{Saha,Wang,Ahmadvand,Maity} The origin of this appears to be lying in the biaxiality of AFM grains and variation in the FM-AFM bias coupling among an ensemble of grains. We report here that we have observed an even more interesting feature of the SEB - variation in the magnitude of the bias depending on the path followed in tracing the hysteresis loop - in a nanocomposite of BiFeO$_3$-Bi$_2$Fe$_4$O$_9$. The hysteresis loop has been traced following two paths - +H$_{max}$ $\rightarrow$ -H$_{max}$ $\rightarrow$ +H$_{max}$ (path a; positive loop) and -H$_{max}$ $\rightarrow$ +H$_{max}$ $\rightarrow$ -H$_{max}$ (path b; negative loop); H$_{max}$ is the maximum field applied for tracing the loop. This $\textit{asymmetry}$ in the SEB offers an additional tunability apart from the magnitude of the maximum field itself and has not been reported for any other composite or multilayer system exhibiting exchange bias. We have also measured the CEB and found that CEB too, exhibits such a path dependency. The asymmetry in the SEB and CEB - $\Delta$H$_{SEB}$ and $\Delta$H$_{CEB}$ - is found to be temperature dependent; while  $\Delta$H$_{SEB}$ decreases with temperature nonmonotonically the $\Delta$H$_{CEB}$ decreases rather monotonically. We have found that the SEB, CEB and their path dependency are originating from a spontaneous breaking of the symmetry of interface magnetic moment and setting of UA among an ensemble of FM and AFM particles  $\textit{even in the absence}$ of first field of hysteresis as a result of superspin glass (SSG) mediated exchange bias coupling interaction. The presence of SSG moment is revealed by a significant memory effect in a stop-and-wait protocol of measurement.\cite{Chen,Sasaki} The memory effect turns out to be depending on the temperature.

\begin{figure}[!h]
  \begin{center}
    \includegraphics[scale=0.25]{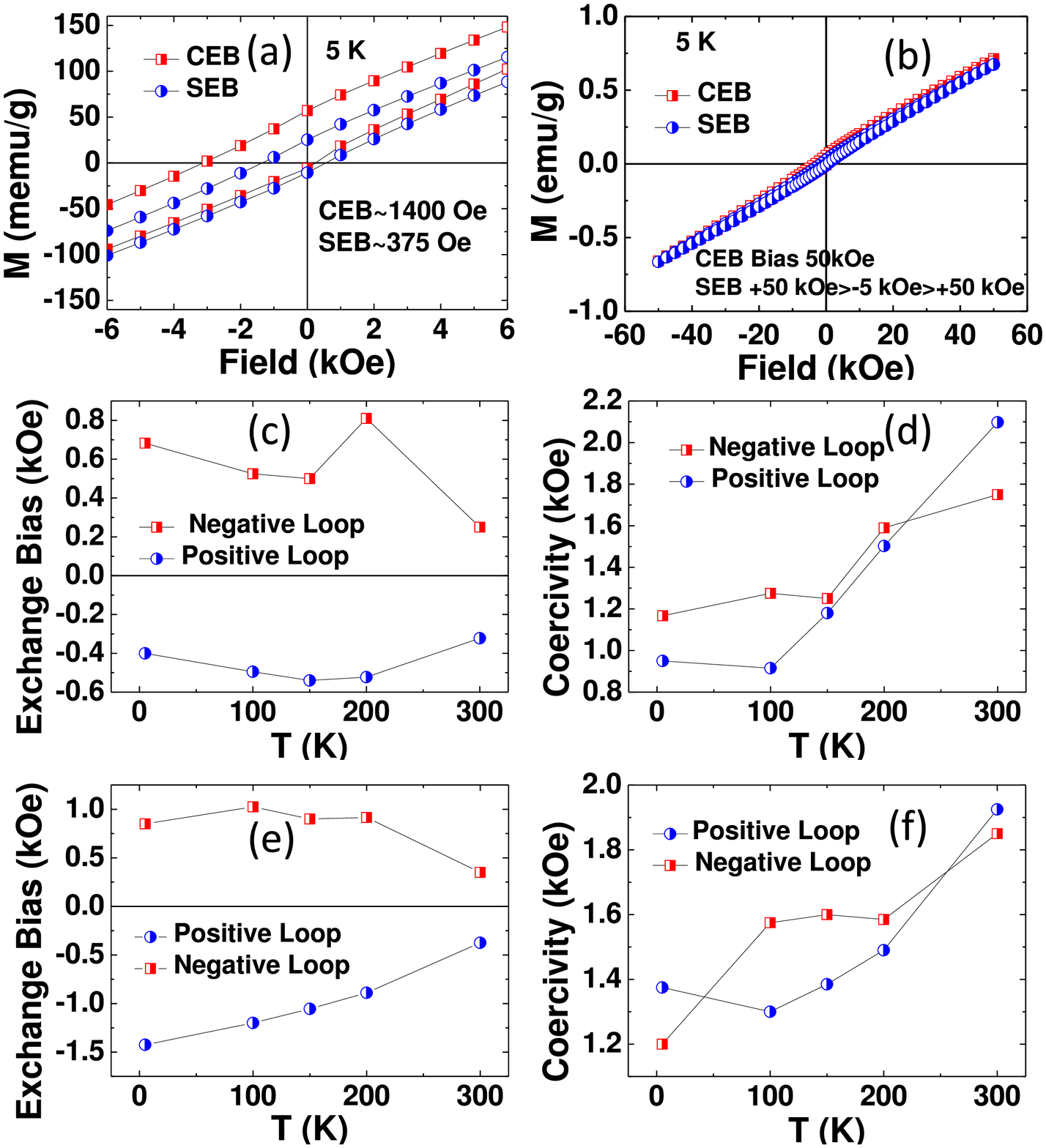} 
    \end{center}
  \caption{(color online) (a) The regions near the origin of the hysteresis loops are blown up to show the SEB and CEB; (b) corresponding full hysteresis loops are shown; the temperature dependence of (c) SEB measured following path-a (positive loop) and path-b (negative loop); and (d) the corresponding coercivity; the temperature dependence of (e) CEB measured following path-a and path-b; and (f) corresponding coercivity. }
\end{figure}

\begin{figure}[!htp]
  \begin{center}
    \includegraphics[scale=0.40]{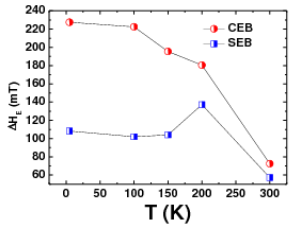} 
    \end{center}
  \caption{(color online) The asymmetry in SEB and CEB - $\Delta$H$_{SEB}$ and $\Delta$H$_{CEB}$ - as a function of temperature.}
\end{figure}

The BiFeO$_3$-Bi$_2$Fe$_4$O$_9$ nanocomposite has been prepared by a solution chemistry route.\cite{Goswami} The volume fraction of the Bi$_2$Fe$_4$O$_9$ phase has been varied from $<$3\% to $\sim$10\%. The exchange bias is maximum for a composite of $\sim$6 vol\% Bi$_2$Fe$_4$O$_9$. It decreases both with the increase and decrease in the volume fraction of the Bi$_2$Fe$_4$O$_9$ phase. The details of the microstructural and crystallographic data and their analyses are available elsewhere.\cite{Maity} The magnetic measurements have been carried out primarily in a SQUID magnetometer (MPMS; Quantum Design) and also in a vibrating sample magnetometer (VSM; LakeShore Cryotronics Inc.). The SEB has been measured following zero-field cooling from $\sim$350 K. In order to ensure that the sample is not biased even by the trapped flux of the superconducting magnet, we have discharged the magnet following appropriate demagnetization protocol where the field is decreased in an oscillation mode. This process leaves a negligible amount of trapped flux - typically ~10 Oe. Before starting a new batch of experiments in MPMS, we normally bring the chamber temperature to 300 K which is above the critical temperature of superconducting coils and the coils are cooled down by filling helium before starting new experiment. We have also measured the SEB following a thermal cycling under zero field through $\sim$800 K, which is far above the magnetic transition points such as blocking temperature T$_B$ ($>$350 K) and Neel point T$_N$ ($\sim$590 K) of the antiferromagnetic component of the composite, in the VSM system for confirming the unbiased state of the sample in MPMS.
  
We report here the results obtained for the nanocomposite with $\sim$6 vol\% Bi$_2$Fe$_4$O$_9$ which exhibits maximum exchange bias. While in an earlier work\cite{Maity} we focussed primarily on the extent of spontaneous and conventional exchange bias observed across a temperature range 5-300 K and path-dependency at $\sim$5 K, we lay more emphasis here how the path dependency of the SEB and CEB evolves with temperature. In Fig. 1a, we show the SEB and CEB by blowing up the portion of the loop near origin. The corresponding full loops are shown in Fig. 1b. Figs. 1c,d,e, and f show the SEB and CEB for a maximum field of 5T measured following two different paths of tracing the loop - path a and path b. Quite clearly both the exchange bias H$_E$ and the coercivity H$_C$ appear to be depending on the path of loop tracing. The sign of the H$_E$ is negative (positive) for positive (negative) starting field. In Fig. 2, we show the asymmetry in the SEB and CEB - $\Delta$H$_{SEB}$ and $\Delta$H$_{CEB}$ - as a function of temperature across 5-300 K. The $\Delta$H$_{SEB}$ exhibits a nonmonotonic pattern with a peak around 200 K. The $\Delta$H$_{CEB}$, of course, decreases with the increase in temperature rather monotonically. We have also measured the CEB using different maximum field H$_m$. In Fig. 3, we show the H$_m$ and temperature dependence of CEB and corresponding H$_C$. Interestingly, while H$_{CEB}$ decreases monotonically with the increase in temperature for different H$_m$ 1, 3, 5T, the corresponding H$_C$ exhibits a rise with temperature from above $\sim$50 K. Therefore, there appears to be an anticorrelation between H$_{CEB}$ and H$_C$. This anticorrelation signifies an anticorrelation between the UA of the exchange coupled structure and the magnetocrystalline anisotropy of the FM component. The magnetocrystalline anisotropy appears to be increasing with the increase in temperature under field cooling. Yet its tensorial nature does not influence the UA of the system. Finally in Fig. 4, we show the signature of the memory effect at different temperatures - measured using a 'stop-and-wait' protocol - within a range below the blocking temperature T$_B$ ($>$350 K) of the system. This has been measured in the following way. The magnetic moment versus temperature pattern is measured initially under zero-field cooling. The temperature is then brought back to 2 K from room temperature under zero field. The magnetic moment versus temperature measurement is then repeated but with a 'stop-and-wait' protocol at a desired temperature. As the desired temperature T$_w$ is reached, the measurement is stopped and waited for a stipulated time - $\sim$10$^4$s. The measurement is then resumed and the temperature is ramped back to room temperature. The difference between the two moment versus temperature plots is shown in Fig. 4. The dip at T$_w$ is the signature of memory effect. It has been shown earlier\cite{Chen,Sasaki} that this memory effect is an unequivocal signature of the presence of SSG moments in the system. However, the dip broadens and the memeory effect weakens as the temperature is raised. This is because of enhanced thermal effect on the spin structure of the system. The thermal energy induces a randomness in the spin structure which, in turn, weakens the memory effect. Interestingly, the memory effect is completely absent above T$_B$. This observation reflects that, as expected, presence of SSG moments and consequent memory effect is conspicuous at only below the T$_B$.

\begin{figure}[!htp]
  \begin{center}
    \includegraphics[scale=0.30]{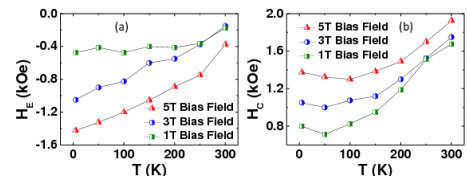} 
    \end{center}
  \caption{(color online) The temperature dependence of (a) conventional exchange bias and (b) corresponding coercivity.}
\end{figure}

\begin{figure}[!htp]
  \begin{center}
    \includegraphics[scale=0.15]{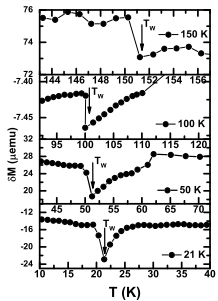} 
    \end{center}
  \caption{The memory effect observed at 21 K, 50 K, 100 K, and 150 K under 'stop-and-wait' protocol of magnetic moment versus temperature measurement. }
\end{figure}

From the detailed analyses of the microstructure and crystallographic data of the nanocomposite,\cite{Maity} it has been found out that the BiFeO$_3$ particles are bigger ($\sim$112 nm) while the Bi$_2$Fe$_4$O$_9$ particles are finer ($\sim$19 nm). It has already been reported by others\cite{Tian} that finer Bi$_2$Fe$_4$O$_9$ particles exhibit FM order. The coarser BiFeO$_3$ particles, on the other hand, are antiferromagnetic with uncompensated local spins. It has also been observed that there are superparamagnetic domains with a blocking temperature T$_B$ $>$350 K.\cite{Maity} The memory effect, on the other hand, signifies the presence of superspin glass moments. As the interparticle distance reduces and the exchange interaction increases, the superparamagnetic domains give way, initially, to superspin glass phase and then even superferromagnetic phase as well.\cite{Chen} Therefore, the spin structure in the BiFeO$_3$-Bi$_2$Fe$_4$O$_9$ composite appears to be consisting of a FM core and SSG shell interacting with the local moments of the AFM structure of coarser BiFeO$_3$ particles. The exchange interaction among the FM cores of different finer particles is considered to have developed a net FM moment across the entire composite. The AFM structures, on the other hand, could be of various types including ones with biaxiality with respect to the axis of application of the field or exchange-coupled pairs.\cite{Saha} This, in turn, yields partially hysteretic, fully hysteretic, and non-hysteretic grains. The symmetry of the interface moment, with respect to the direction of applied field, within the ensemble of coupled grains is spontaneously broken $\textit{even in the absence of first field of hysteresis loop tracing}$ via an indirect exchange bias coupling interaction between FM core of Bi$_2$Fe$_4$O$_9$ and AFM moments of BiFeO$_3$ through the intermediate SSG moments at the interface. This spontaneous setting of UA along the negative direction of applied field (or universal UA) even under zero field is the origin of the path dependency for both SEB and CEB. The volume fraction of the partially hysteretic grains v$^{UUA}_{fg}$ with universal UA (UUA) along the negative direction of the applied field governs the magnitude of exchange bias as well as its path dependency. The volume fraction of the partially hysteretic grains v$^{UA}_{fg}$ with UA set by the first field of the loop tracing, on the other hand, governs the magnitude of exchange bias but not the path dependency. The temperature dependences of v$^{UUA}_{fg}$(T) and v$^{UA}_{fg}$(T) and their subtle interplay influence the temperature dependence of $\Delta$H$_{SEB}$ and H$_{SEB}$. The nonmonotonicity in both H$_{SEB}$ and $\Delta$H$_{SEB}$ possibly results from an initial increase in v$^{UUA}_{fg}$ and v$^{UA}_{fg}$ with temperature due to an increase in SSG mediated indirect exchange bias coupling among the grains. As the temperature increases, the frozen moments of SSG at the shell in between FM and AFM grains are thermally activated to interact strongly with the FM and AFM moments. This strong interaction, in turn, makes the spontaneous symmetry breaking more effective and gives rise to enhanced path dependency in SEB. With further rise in temperature, the v$^{UUA}_{fg}$ eventually decreases as enhanced thermal randomization of the spin structure itself results in weakening of bias coupling interaction. This anomalous influence of temperature is not conspicuous in the case of CEB as in this case both the UUA and UA are further influenced by field cooling from higher temperature. The impact of field cooling masks the subtle role of temperature on SSG induced spontaneous setting of UA. In fact, as shown in Fig. 3, apart from its path dependency, the CEB itself does not exhibit any nonmonotonicity across 5-300 K.  

The $\Delta$H$_{SEB}$ and $\Delta$H$_{CEB}$ offer an additional tunability to the exchange bias. Using a combination of maximum field of loop tracing (H$_m$) as well as the path followed in tracing the loop - positive or negative - it is possible to tune the magnitude of the exchange bias. This tunability, in turn, can increase the functionality in electrically switching the magnetic anisotropy of a ferromagnetic system in a BiFeO$_3$-ferromagnetic composite like the present one via multiferroic coupling between ferroelectric polarization and magnetization in BiFeO$_3$. 

In summary, we report that in a nanocomposite of ($\sim$94 vol \%)BiFeO$_3$-($\sim$6 vol\%)Bi$_2$Fe$_4$O$_9$ with finer and ferromagnetic Bi$_2$Fe$_4$O$_9$ particles and coarser and antiferromagnetic BiFeO$_3$, one observes a large and path-dependent spontaneous exchange bias ($\sim$30-60 mT) across 5-300 K. The conventional exchange bias too is found to be path dependent. This path dependency offers an additional tunability in the effect of electrical switching of magnetic anisotropy in a BiFeO$_3$-ferromagnetic composite via multiferroic coupling and is expected to improve the functionality of such a device enormously.

This work has been supported by Indo-Ireland joint program (DST/INT/IRE/P-15/11), Science Foundation of Ireland (SFI) Principal Investigator (PI) Project No. 11/PI/1201, and FORME SRC project (07/SRC/I1172) of SFI. One of the authors (S.G.) acknowledges support from a Research Associateship of CSIR.

\end{document}